# Investigation of anisotropic metric in multi-dimensional space-time and generation of additional dimensions by scalar field

Sergey V.Yakovlev
sergey.y@mail.ru

Moscow, January 20, 2011

Were investigated anisotropic metric of higher dimensional space-time with cosmological term and scalar field. Showed, that presence of scalar field is equivalent to anisotropic metric in the multidimensional space-time and proposed idea of dimensions generation by scalar field. Were solved Einstein's equations for higher dimensional space-time of Kazner type and derived expressions for density of energy for scalar field, which generate additional dimensions, and proposed the procedure of renormalization of the metric.

As the main model of anisotropic space-time we'll consider Kazner model. Let's recall for the beginning the main facts about that metric. Kazner metric (1922) is a private matter of using of Gauss normal coordinates, which we could introduce by the next manner. In any space-time, not obligatory isotropic, we'll take initial space-like surface $S_1$, with coordinates $(x^1, x^2, x^3)$. From the points of that space-like surface we'll start the orthogonal to surface geodesic world lines and parametrize them by such a way, that $(x^1, x^2, x^3) = const, x^0 = t = t_1 + \tau$, where $\tau$ - time along each world line, and $S_1$ is related to the moment $\tau = 0$. In such coordinate system, that is called normal Gaussian, metric is getting the following form:

$$ds^2 = -dt^2 + g_{ij}(t)dx^i dx^j, \qquad (1)$$

because $g_{oi}(0) = 0$, and it could be shown that $g_{oi}(t) = 0$, where we don't have dependence from time $t$. That's because the vectors $\vec{e}_0$ are tangent to normal curves of surfaces, and they commutate with vectors of coordinate basis $\vec{e}_k$ located on tangent surface:

$$\frac{d(\vec{e}_0 \vec{e}_i)}{d\tau} = \nabla_{\vec{e}_0}(\vec{e}_0 \vec{e}_i) = \vec{e}_0 \nabla_{\vec{e}_0} \vec{e}_i = \vec{e}_0 \nabla_{e_i} \vec{e}_0 = \frac{1}{2}\nabla_{e_i}(\vec{e}_0 \vec{e}_0) = 0,$$

$$[\vec{e}_0, \vec{e}_i] = 0, \quad \vec{e}_0 = \frac{\partial}{\partial \tau}. \quad i = 0,...,n.$$

For obtaining Kazner metric let's consider anisotropic four-dimensional model with metric (1), where space-like sectors $t = const$ have flat geometry. In gravitational dominated case, when energy-momentum tensor $T_{\mu\nu} = 0$, by calculating Ricci tensor $R_{\mu\nu}$ and equaling it to zero, we obtain

$$g_{ij}(t) = \delta_{ij} t^{2P_i}$$

Finished expression for the metric, called Kazner metric, is

$$ds^2 = -dt^2 + t^{2P_i} dx^{i^2} \qquad (2)$$

where constants $P_i$ should satisfy the following equations

$$\sum_i P_i = 1, \quad \sum_i P_i^2 = 1 . \qquad (3)$$

The volume of such Universe is proportional to time $t$:

$$\sqrt{g} = t, \quad g = \det g_{ij}(t).$$

But that is anisotropic Universe. Lengths, that are parallel to different coordinate axes, is increasing with different velocities, and along with the one of that axes length could be diminishing, not increasing.

Mathematically we could see contracting from the fact, that in eq.(3) for $P_i$ one of the constants, for example $P_3$, should be negative (in the case of three dimensional space-like surface):

$$-\frac{1}{3} \leq P_3 \leq 1; \quad P_1, P_2 \geq 0 .$$

Accordingly, along that direction we would see the blue shift of radiation instead of the red one in the reality. Thus, in case of three-dimensional space Kazner model is not real for isotropic expanding of the Universe at least for the latest time of expanding.

That difficulty could be solved, if we'll introduce additional dimensions. And all anisotropy of Kazner model of the Universe we'll put for extra-dimensions, while our three-dimensional space is expanding isotropic .

For the simplest case, when the number of additional dimensions is $d$, and the number of all dimensions is equal $n = 3 + d$, and they are isotropic, we could write down

$$P_1 = P_2 = P_3 = P, \quad P_4 = P_5 = ... = P_d = Q,$$

and according the conditions (3) for $P$ and $Q$ we have

$$3P + (n-3)Q = 1, \quad 3P^2 + (n-3)Q^2 = 1 .$$

Solving the system of that equations, we obtain

$$P = \frac{1}{n}\left(1 \pm \sqrt{\frac{(n-1)(n-3)}{3}}\right)$$

$$Q = \frac{1}{n}\left(1 \mp \sqrt{\frac{3(n-1)}{(n-3)}}\right)$$

or, if we mean $P > 0, Q < 0$, we have

$$P = \frac{1}{n}\left(1 + \sqrt{\frac{(n-1)(n-3)}{3}}\right)$$

$$Q = \frac{1}{n}\left(1 - \sqrt{\frac{3(n-1)}{(n-3)}}\right) \qquad (4)$$

Metric is getting the form

$$ds^2 = -dt^2 + t^{2P}(dx^{1^2} + dx^{2^2} + dx^{2}) + t^{2Q}(dy^{1^2} + dy^{2^2} + ... + dy^{d^2}),$$

where $y^i, i = 1,...,d$ is additional dimensions. For example if $d = 1, n = 4$, using (4) we have

$$P = \frac{1}{2}, \quad Q = -\frac{1}{2}.$$

We see, that $d$ - dimensional space-like surface $M^d$ contracts in isotropic way. Taking the number of dimensions equals to infinity, we have

$$\lim_{n \to \infty} P(n) = \frac{1}{\sqrt{3}}, \quad \lim_{n \to \infty} Q(n) = 0,$$

- space is expanding only along the three dimensions, the rest of them remains constant in time.

By such simple mechanism we could explain unobserved for us extra-dimensions with isotropic submanifolds in Kazner models. In case of infinite dimensional space, after taking a limit, Kazner conditions are not more valid.

Let's make generalization of the above model, saving specific conditions (3) for $P_i$. Parametrizing space-like metric in next manner [5]

$$g_{ij}(t) = e^{2\alpha(t)} e^{2\beta_{ij}(t)}, \tag{5}$$

where $\beta_{ij}$ - symmetric matrix without trace, we have

$$\sqrt{g} = e^{n\alpha}.$$

Define the matrix

$$P_{ij} = d(\ln g)_{ij} / d \ln g.$$

Then we obtain

$$P_{ij} = \frac{1}{n}\left(\delta_{ij} + \frac{d\beta_{ij}}{d\alpha}\right).$$

Kazner conditions are getting the form

$$Sp\beta_{ij} = 0,$$
$$\sum_{i,j}\left(\frac{d\beta_{ij}}{d\alpha}\right)^2 = n(n-1). \tag{6}$$

In the case of diagonal metric we have

$$g_{ij}(t) = \delta_{ij} e^{2\alpha(t)} e^{2\beta_{ij}(t)}. \tag{7}$$

Introducing orthonormal basic 1-form $\omega^i = e^{\alpha+\beta_i} dx^i$, we could get by standard way the non-zero forms of curvature

$$R^i_j = \dot{\gamma}_i \dot{\gamma}_j \omega^i \wedge \omega^j, \quad R^i_0 = (\ddot{\gamma}_i + \dot{\gamma}_i^2)\omega^0 \wedge \omega^i,$$

and get the Einstein equations

$$\left(\sum \dot{\gamma}_i\right)^2 - \sum \left(\dot{\gamma}_i\right)^2 = 16\pi T_{00},$$

$$\ddot{\gamma}_j + \dot{\gamma}_j \sum \dot{\gamma}_i - \frac{1}{n}\left(\sum \dot{\gamma}_i\right)^2 - \frac{1}{n}\sum \ddot{\gamma}_j = 8\pi\left(T_{jj} - \frac{1}{n}T^k_k\right),$$

$$\gamma_i = \alpha + \beta_i.$$

Using the first Kazner condition $Sp\beta_{ij} = 0$ gives us the system of equation:

$$\dot{\alpha}^2 = \frac{16\pi}{n(n-1)}\left(T_{00} + \frac{1}{16\pi}\sum_i \dot{\beta}_i^2\right)$$

$$\ddot{\beta}_j + n\dot{\alpha}\dot{\beta}_j = 8\pi\left(T_{jj} - \frac{1}{n}T^k_k\right) \quad (8)$$

Components of energy-momentum tensor here relate to orthonormal basis $\omega^i = e^{\alpha+\beta_i}dx^i$.
The second Kazner condition results from the first Einstein equation, because of the $T_{00} = 0$, or in the rough

$$T_{00} \ll \frac{c^2}{16\pi G}\sum_i \dot{\beta}_i^2 = \rho_{anis}, \quad (9)$$

where $\rho_{anis}$ is called effective anisotropy density of energy, $G = \frac{1}{M^2}$ - gravitational constant for $(n+1)$ space-time, $M$ - Planck mass in manifold $M^{n+1}$, $c$ – velocity of light (here we're getting back to original system of value).

Presence of the matter breaks the second Kazner condition. Let's consider the case of empty space time with $n$ space dimensions and non-zero's cosmological constant $\Lambda$.
Einstein system (7) takes the form

$$\dot{\alpha}^2 = \frac{1}{n(n-1)}\left(2\Lambda + \sum_i \dot{\beta}_i^2\right)$$

$$\ddot{\beta}_j + n\dot{\alpha}\dot{\beta}_j = 0 \quad (10)$$

Here we used the second Kazner condition (6). Further we'll consider full isotropic space-like submanifolds $M^3$ and $M^d$, $M^n = M^3 \otimes M^d$, where $d$ is number of additional dimensions, so first equation in (6) is turning into following equation

$$3\beta + d\beta_d = 0, \quad (11)$$

where $\beta$ is related to $M^3$, and $\beta_d$ to $M^d$. From the second eq.(8) we get for $\beta_k$:

$$\beta_k = \beta_{0_k} + \frac{B_k}{B}\sqrt{\frac{n-1}{n}}\left[n\alpha - \ln\left(1+\sqrt{1+\frac{2\Lambda}{B^2}e^{2n\alpha}}\right)\right], \tag{12}$$

where $\beta_0, B_k$ -constants of integration,

$$\dot{\beta}_k = B_k e^{-n\alpha}, \quad B^2 = \sum_{k=1}^{n} B_k^2, \quad \sum_{k=1}^{n} B_k = 0, \quad \sum_{k=1}^{n} \beta_{0k} = 0.$$

From the system (10), we have finally

$$\alpha = \frac{1}{n}\ln\left[\frac{B}{\sqrt{2\Lambda}}sh(\lambda(t-t_0)+\delta)\right], \quad \beta_i = C_i \ln(\frac{B}{\sqrt{2\Lambda}}th((\lambda(t-t_0)+\delta)/2)), \quad \Lambda > 0;$$

$$\alpha = \frac{1}{n}\ln\left[\frac{B}{\sqrt{2|\Lambda|}}Sin\lambda t\right], \quad \beta_i = C_i \ln(\frac{B}{\sqrt{2|\Lambda|}}tg(\lambda t/2)), \quad \Lambda < 0;$$

(13)

$$\alpha = \frac{1}{n}\ln\left[\sqrt{\frac{n}{n-1}}Bt\right], \quad \beta_i = C_i \ln\left[\sqrt{\frac{n}{n-1}}Bt\right], \quad \Lambda = 0;$$

$$C_i = \frac{B_i}{B}\sqrt{\frac{n-1}{n}}, \quad \lambda = \sqrt{\frac{2|\Lambda|n}{n-1}}, \quad e^\delta = \frac{\sqrt{2\Lambda}}{B}.$$

When $\Lambda > 0$, we could impose initial conditions on $\alpha$ when $t=0$: the first one is $\lambda t_0 = \delta$, and second one is $t_0 = 0$, $\delta = e^\alpha = 0$, that is

$$2\Lambda = B^2.$$

Second case is interested for clarifying nature of the gravitational constant $\Lambda$, which could depends on the initial dates describing dynamic of space dimensions in the model.

Respectively for the time-dependant scale factors in isotropic submanifolds $M^3$ and $M^d$, using eq.(11) and introducing independent on time terms into the new coordinates (rescaling coordinates), we get

$$r(t) \equiv e^{\alpha+\beta} = \left[sh\lambda t \left(th(\lambda t/2)\right)^{\sqrt{\frac{(n-1)(n-3)}{3}}}\right]^{\frac{1}{n}},$$

(14)

$$r_d(t) \equiv e^{\alpha+\beta_d} = \left[sh\lambda t \left(th(\lambda t/2)\right)^{-\sqrt{\frac{3(n-1)}{(n-3)}}}\right]^{\frac{1}{n}}, \quad \Lambda > 0;$$

$$r(t) \equiv e^{\alpha+\beta} = \left[ Sin\lambda t \left(tg(\lambda t/2)\right)^{\sqrt{\frac{(n-1)(n-3)}{3}}} \right]^{\frac{1}{n}},$$

$$r_d(t) \equiv e^{\alpha+\beta_d} = \left[ Sin\lambda t \left(tg(\lambda t/2)\right)^{-\sqrt{\frac{3(n-1)}{(n-3)}}} \right]^{\frac{1}{n}}, \quad \Lambda < 0.$$

(15)

From (14) we could see that when $t \to \infty$, $r = r_d$ (surely with accuracy up to the constant). Decision with $\Lambda < 0$ is describing circled Universe with recollaps. For different $t, n$, ratios between scale factors in $M^3$ and $M^d$ are changing in different limits.

`   Let's obtain the metric for infinite dimensional space. Not considering physical sense and correctness of the limit in model, by putting $n \to \infty$, we get

$$r(t) = \left(th\sqrt{\Lambda/2}t\right)^{\sqrt{\frac{1}{3}}}, \quad r_d(t) = 1, \quad \Lambda > 0 \; ;$$

$$r(t) = \left(tg\sqrt{|\Lambda|/2}t\right)^{\sqrt{\frac{1}{3}}}, \quad r_d(t) = 1, \quad \Lambda < 0.$$

Let's now fill up the Universe with uniform scalar field $\phi$ with Lagrangian

$$L = -\frac{1}{8\pi}\phi_{,\alpha}\phi^{,\alpha}$$

Energy-momentum tensor for the field is following

$$T_{\mu\nu} = \frac{1}{4\pi}\left(\phi_{,\mu}\phi_{,\nu} - \frac{1}{2}g_{\mu\nu}\phi_{,\alpha}\phi^{,\alpha}\right),$$

with zero's component

$$T_{00} = \frac{1}{8\pi}\dot{\phi}^2.$$

Klein-Gordon equation for the metric (5)

$$\left(\phi^{,\alpha}(-g)^{1/2}\right)_{,\alpha} = \ddot{\phi} + n\dot{\alpha}\dot{\phi} = 0.$$

(16)

System of the Einstein equations (8) is getting the form

$$n(n-1)\dot{\alpha}^2 = 2\Lambda + 2\dot{\phi}^2 + \sum \dot{\beta}_i^2,$$

$$\ddot{\beta}_i + n\dot{\alpha}\dot{\beta}_i = 0.$$

(17)

From equations (16) and (17) we could make suggestion, that if we renormalize constants $\alpha, \beta_i$, then they will correspond the describing of the new metric in space with more dimensions, at least of 1 or any number of space dimensions, because the Klein-Gordon equation exactly coincide with the second Einstein equation (16).

The presence of massless scalar field in that case leads to appearance, or generation of new space-like dimension. Such reformulation of the task is possible because of the coincident of the shape of equation. Let's renormalize the constants in the following way:

$$\alpha_N = A\alpha, \quad \dot{\sigma}_{i_i} = b_i \dot{\beta}_i, \quad i = 1,...,N..$$

The Einstein equations in $M^N = M^3 \otimes M^D$ have the following form:

$$N(N-1)\dot{\alpha}_N^2 = 2\Lambda_N + \sum_{i=1}^{N} \dot{\sigma}_i^2,$$  (18)

$$\ddot{\sigma}_i + N\dot{\alpha}_N \dot{\sigma}_i = 0,$$

So we've got $A = \dfrac{n}{N}$ and next equation for normalization coefficients,

$$\dot{\alpha}^2 = \frac{3\dot{\beta}^2}{n(n-1)} + \frac{(n-3)\dot{\beta}_d^2}{n(n-1)} + \frac{2\dot{\phi}^2}{n(n-1)} + \frac{2\Lambda}{n(n-1)} =$$

$$= \frac{3N\dot{\sigma}^2}{n^2(N-1)} + \frac{(n-3)N\dot{\sigma}_d^2}{n^2(N-1)} + \frac{(N-n)N\dot{\sigma}_D^2}{n^2(N-1)} + \frac{2N\Lambda_N}{n^2(N-1)},$$

where $\beta$ and $\sigma$ is related to $M^3$, $\beta_d$ to $M^{n-3}$, $\sigma_d$ and $\sigma_D$ to $M^{N-3}$.

From here it's easy to find out the conditions for renormalization of metric in $M^n$ and $M^N$:

$$\dot{\beta}^2 = \frac{(n-1)}{n}\frac{N}{N-1}\dot{\sigma}^2, \qquad \dot{\beta}_d^2 = \frac{(n-1)}{n}\frac{N}{(N-1)}\dot{\sigma}_d^2$$

$$\dot{\phi}^2 = \frac{(n-1)}{2n}\frac{(N-n)N}{(N-1)}\dot{\sigma}_D^2, \qquad \Lambda_N = \frac{n}{n-1}\frac{N-1}{N}\Lambda,$$  (19)

where solutions for $\sigma_k$ have the form (12), (13). One scalar field $\phi$ generates $N-n$ additional dimensions, which are describing by the new rescaled field $\sigma_D$. The most simple case is generation of only one dimension by one massless scalar field, that is $N = n+1$.

From the form of energy-momentum tensor for massless scalar field, we would introduce in general $n$ massless scalar fields $\theta_i$ for isotropic submanifold $M^3$

$$\theta_i = \sqrt{3}\beta_i, \quad \theta = \theta_1 = \theta_2 = \theta_3,$$

and $d$ is equal the number scalar fields $\theta_d$ for isotropic submanifold $M^d$

$$\theta_d = \sqrt{3}\beta_d, \quad \theta_d = \theta_4 = ... = \theta_n;$$

$\sqrt{3}$ is introduced for right normalizing of the field in the right side of Einstein equations, that should correspond eq.(10).

In general, such an approach for Kazner metric (7) gives us that massless scalar fields could generate metric in manifold and dimensions. Zero's component of energy-momentum tensor of scalar field, that is additional term to the vacuum energy $\Lambda$, for the fields $\theta_i$ has the form

$$T_{00}(\theta) = \frac{1}{16\pi}(3\dot{\theta}^2 + d\dot{\theta}_d^2).$$

Using the first of Einstein eq.(10), or that is the same, broken 2-nd Kazner condition, we obtain

$$T_{00}(\alpha) = \frac{n(n-1)}{16\pi}\dot{\alpha}^2 - \frac{\Lambda}{8\pi} = \frac{B^2}{16\pi}e^{-2n\alpha},$$

$$T_{00}(t) = \frac{\Lambda}{8\pi sh^2 \lambda t}, \quad \Lambda > 0$$

$$T_{00}(t) = \frac{\Lambda}{8\pi Sin^2 \lambda t}, \quad \Lambda < 0$$

$$T_{00}(t) = \frac{n-1}{16\pi n t^2}, \quad \Lambda = 0.$$

- the density of energy is decreasing with time, where for the start moment we've took $t = 0$.

Thus, by considered above way, existence of extra-dimensions in the model leads to appearance in right side of Einstein's equations massless scalar fields and on the contrary.

Choice of Kazner type metric could be justified cause such metric describes flat space. We're introducing massless scalar fields and obtaining solutions with cosmological term in the manifold with more number of space dimensions and without presence of matter. Non-zero choice of cosmological constant in fact corresponds the weak break of the second Kazner condition, and for sure is a little, but dominate factor for investigating of metric. Because $\Lambda$ is little, and when equation (9) for energy of isotropism is valid, we could consider right the first evaluation of standard Kazner solution (4).

Of course, applying the decision around singularity is not valid in general case. Singularity in such model is just consequence of using the weak mathematical approach, than the real physics situation.

Postulating the high degrees of symmetry, for example space continuous, is necessary for detailed mathematical analysis [4]. With presence of high symmetry we have got necessarily that at some moment of time matter in the Universe must concentrate in point or in line or two-dimensional surface and on, that in general corresponds to submanifold $M^3$ as $M^d$.